\documentclass[reprint,superscriptaddress,showpacs]{revtex4-1}
\usepackage{amsmath}
\usepackage{graphicx}
\usepackage{amssymb}

\begin{document}

\title{Steady-state squeezing and entanglement in a dissipatively coupled NOPO network}

\author{Yoshitaka Inui}
\email[]{inui@nii.ac.jp}

\affiliation{NTT Basic Research Laboratories, Morinosato 3-1, Atsugi, Kanagawa 243-0198, Japan}
\affiliation{National Institute of Informatics, Hitotsubashi 2-1-2, Chiyoda-ku, Tokyo 101-8430, Japan}

\author{Yoshihisa Yamamoto}

\affiliation{NTT Basic Research Laboratories, Morinosato 3-1, Atsugi, Kanagawa 243-0198, Japan}
\affiliation{E. L. Ginzton Laboratory, Stanford University, Stanford, CA 94305, USA}

\date{\today}

\begin{abstract}

We investigate the steady-state photon-number squeezing and quantum entanglement 
in a network of nondegenerate optical parametric oscillators (NOPOs). 
We treat each NOPO with Shen's Raman laser model, 
whose lasing mode provides a photon-number-squeezed state. 
Two dissipatively coupled NOPOs satisfy Hillery-Zubairy's $HZ1$ entanglement criterion 
if they are pumped far above the threshold and the dissipative coupling is 
sufficiently larger than the NOPO cavity loss. 

\end{abstract}


\maketitle

\section{Introduction}

Gain-dissipative analogs of closed-space equilibrium spin systems, 
such as Ising, XY, and Heisenberg models, 
have been studied in laser and parametric oscillator networks 
\cite{Eckhouse08,Nixon13,Pal17,Tamate16,Hamerly16,Takeda17,Wang13,Takata15,Maruo16,Marandi14,Takata16,Inagaki16}. 
The presence of dissipative coupling between oscillators creates an ordered state, 
which is similar to the dissipatively ordered atomic polarizations in superradiance \cite{Gross82,Temnov09}. 
When each laser has $U(1)$ continuous degrees of freedom due to 
the lack of phase restoring potential \cite{Hamerly16}, 
a gain-dissipative XY model is obtained by putting linear couplings between the lasers 
\cite{Eckhouse08,Nixon13,Pal17,Tamate16,Hamerly16,Takeda17}. 
Similarly, degenerate optical parametric oscillators (DOPOs) with $Z_2$ degrees of freedom \cite{Hamerly16} 
have yielded gain-dissipative Ising models \cite{Wang13,Takata15,Maruo16,Marandi14,Takata16,Inagaki16}. 
These systems are used for solving combinatorial optimization problems \cite{Marandi14} 
and simulating equilibrium spin models \cite{Pal17,Tamate16,Takeda17,Inagaki16}. 
The two-dimensional classical XY model was shown to possess 
a Berezinskii-Kosterlitz-Thouless (BKT) phase transition 
whose low-temperature phase has a quasi-long-range order \cite{Berezinskii71,Kosterlitz73}. 
A two-dimensional lattice of gain-dissipative XY model is also expected to have such a phase \cite{Tamate16,Pal17}. 

The classical XY model is represented by nearest-neighbor-coupled coherent spin states \cite{Arecchi72} 
on the equatorial plane of a Bloch sphere, or by coherent boson states \cite{Glauber63}. 
However, some related models showing the BKT phase are modified by quantum fluctuations. 
The quantum XY model \cite{Lieb61,Katsura62} is represented by quantum spin operators. 
As a result of quantum fluctuation, the ground state energy the quantum XY model has, 
is different from that in the classical XY model\cite{Lieb61,Katsura62}. 
Dissipative laser network with simple coherent states simulates only the classical XY model. 
When we use nonclassical lasing states, however, 
the impact of quantum fluctuations will be introduced into the laser network. 
We consider the photon-number-squeezed state, predicted in 
a pump-noise-suppressed laser\cite{Yamamoto86}, a Raman laser\cite{Ritsch92,Gheri92}, 
and a single-quantum-dot laser\cite{Benson99}, as a building block of network. 
The photon-number-squeezed lasers provide a second-order 
intensity correlation function smaller than $g^{(2)}(0)=1$ for coherent state, 
and will have larger phase fluctuations due to the uncertainty principle. 

In this paper, we show that the photon-number-squeezed state and 
the entangled state exist in a system of two coupled 
nondegenerate optical parametric oscillators (NOPOs) \cite{Hamerly16,Takeda17}. 
As the model of an NOPO, we use the Shen's Raman laser model\cite{Shen67}. 
In an NOPO with second-order optical nonlinearity $\chi^{(2)}$, 
when the idler mode is treated as an environment, 
an NOPO is equivalent to the Shen's model where electronic excitations are treated as an environment. 
In an early quantum treatment of an NOPO, where the signal and the idler mode have the same dissipation rates, 
the photon number squeezed state with $g^{(2)}(0)<1$ was 
theoretically obtained for the signal mode above the threshold \cite{McNeil83}. 
As shown by their work, even after tracing out the idler mode, 
the signal mode of an NOPO above the threshold can have photon-number-squeezed state. 
The steady-state photon-number-squeezing was also predicted \cite{Ritsch92,Gheri92} 
in a coherently excited three-level laser system. 
Above the threshold, they showed that Mandel's $Q$ parameter\cite{Mandel79} converges as $Q\rightarrow -1/2$. 
This is smaller than $Q\rightarrow -1/4$ of the symmetric NOPO 
where the signal and the idler modes have the same loss \cite{McNeil83}. 
The spectral analysis was performed for symmetric NOPO \cite{Bjork88}. 
It is shown that above the threshold, the signal mode shows sub-shot-noise spectral intensity below the cavity cut-off frequency. 
Similar spectral analysis was also carried out \cite{Roos03}, 
in the limit where the electronic excited states have much larger dissipation than the signal mode. 

In this paper, for optical components of the coherent XY machine, 
we only consider an NOPO described by Shen's model\cite{Shen67} due to the theoretical simplicity, 
although hyperparametric oscillation with third-order nonlinear susceptibility $\chi^{(3)}$ 
has been experimentally achieved \cite{Takeda17}. 
We first show that Shen's model has photon number squeezing above the threshold. 
It is known that, for the dissipatively coupled DOPOs\cite{Marandi14}, where the canonical momentum of the signal mode is squeezed, 
an entangled state is theoretically predicted even above the threshold, via dissipative linear coupling \cite{Takata15,Maruo16}. 
Similarly, dissipatively coupled photon-number-squeezed states also attain entangled states. 
This paper is organized as follows. 
In Sec.II, we present the NOPO model and derive the density matrix master equation of the dissipatively coupled two NOPOs. 
In Sec.III, we introduce analytical and numerical methods. 
We show that the positive-$P$ $c$-number stochastic differential equation (PSDE) \cite{Takata15} 
and truncated Wigner stochastic differential equation (WSDE) \cite{Maruo16} 
predict the photon-number squeezing with excitation of more than twice the oscillation threshold. 
In Sec.IV, we numerically study a system of two NOPOs with dissipative coupling 
and show that it can attain entangled states, 
if each NOPO operates far above the threshold and the dissipative coupling is sufficiently large. 
In Sec.V, we present the summary of this paper. 

\section{Quantum Master Equation}

In this section, we introduce the density matrix master equation for ferromagnetically coupled NOPOs. 
The interaction between the intra-cavity pump mode $\hat{a}_p$, signal mode $\hat{a}_s$, and idler mode $\hat{a}_i$ is represented as
\begin{equation}
\hat{H}_{NL}=i\hbar \kappa (\hat{a}_s^{\dagger}\hat{a}_i^{\dagger}\hat{a}_p-\hat{a}_p^{\dagger}\hat{a}_i\hat{a}_s). 
\end{equation}
The quantum master equation of a single NOPO is as follows: 
\begin{equation}
\label{sNOPO}
\frac{\partial \hat{\rho}}{\partial t}=-\frac{i}{\hbar}[\hat{H}_{NL}+H_{exc},\hat{\rho}]+\sum_{k=p,s,i}\gamma_k ([\hat{a}_k,\hat{\rho}\hat{a}_k^{\dagger}]+[\hat{a}_k\hat{\rho},\hat{a}_k^{\dagger}])
\end{equation}
where $[\hat{X},\hat{Y}]=\hat{X}\hat{Y}-\hat{Y}\hat{X}$. 
The coherent excitation of the intra-cavity pump mode is represented by the following Hamiltonian: 
\begin{equation}
\hat{H}_{exc}= i \hbar \varepsilon ( \hat{a}_{p}^{\dagger} - \hat{a}_{p} ) . 
\end{equation}
Here, we neglect the detunings in both the NOPO interaction and in the excitation.  
The cavity linewidths (half-width at half maximum) $\gamma_k(k=p,s,i)$ represent 
the dissipation of the pump, signal and idler mode, respectively. 

We assume a large dissipation rate of the idler mode. 
We can adiabatically eliminate the idler mode by using the Lindblad procedure, 
assuming that the photon number in the idler mode is negligible due to rapid decay of the idler mode. 
We obtain the following Liouvillian: 
\begin{equation}
\mathcal{L}_{NL}\hat{\rho}=G ([\hat{a}_s^{\dagger}\hat{a}_p,\hat{\rho}\hat{a}_p^{\dagger}\hat{a}_s]+[\hat{a}_s^{\dagger}\hat{a}_p\hat{\rho},\hat{a}_p^{\dagger}\hat{a}_s])
\end{equation}
This Liouvillian has been obtained by eliminating the fermionic modes in the Raman scattering 
with the $\Lambda$ configuration\cite{Shen67}, 
and also obtained in the Raman scattering with bosonic excitation \cite{Walls73,McNeil74}. 
If the idler mode has the large linewidth ($\gamma_i\gg \gamma_p,\gamma_s$), 
$G$ is represented as $G=\kappa^2/\gamma_i$. 
After such elimination of the idler mode, the master equation of a single NOPO $\frac{\partial \hat{\rho}}{\partial t}=\mathcal{L}_{NOPO}\hat{\rho}$ is as follows: 
\begin{equation}
\label{qme1}
\mathcal{L}_{NOPO}\hat{\rho}=-\frac{i}{\hbar}[\hat{H}_{exc},\hat{\rho}]+\sum_{k=A,B,C}([\hat{L}_k,\hat{\rho}\hat{L}_k^{\dagger}]+{\rm h.c.})
\end{equation}
where $\hat{L}_k(k=A,B,C)$ represent Liouvillian terms of a single NOPO. 
Here, $\hat{L}_A=\sqrt{\gamma_p}\hat{a}_{p}$ and $\hat{L}_B=\sqrt{\gamma_s}\hat{a}_{s}$, and 
$\hat{L}_C$ represents spontaneous and stimulated parametric scattering from the pump mode to the signal mode: $\hat{L}_{C}=\sqrt{G}\hat{a}_{s}^{\dagger}\hat{a}_{p}$. 

\begin{figure}
\begin{center}
\includegraphics[width=8.0cm]{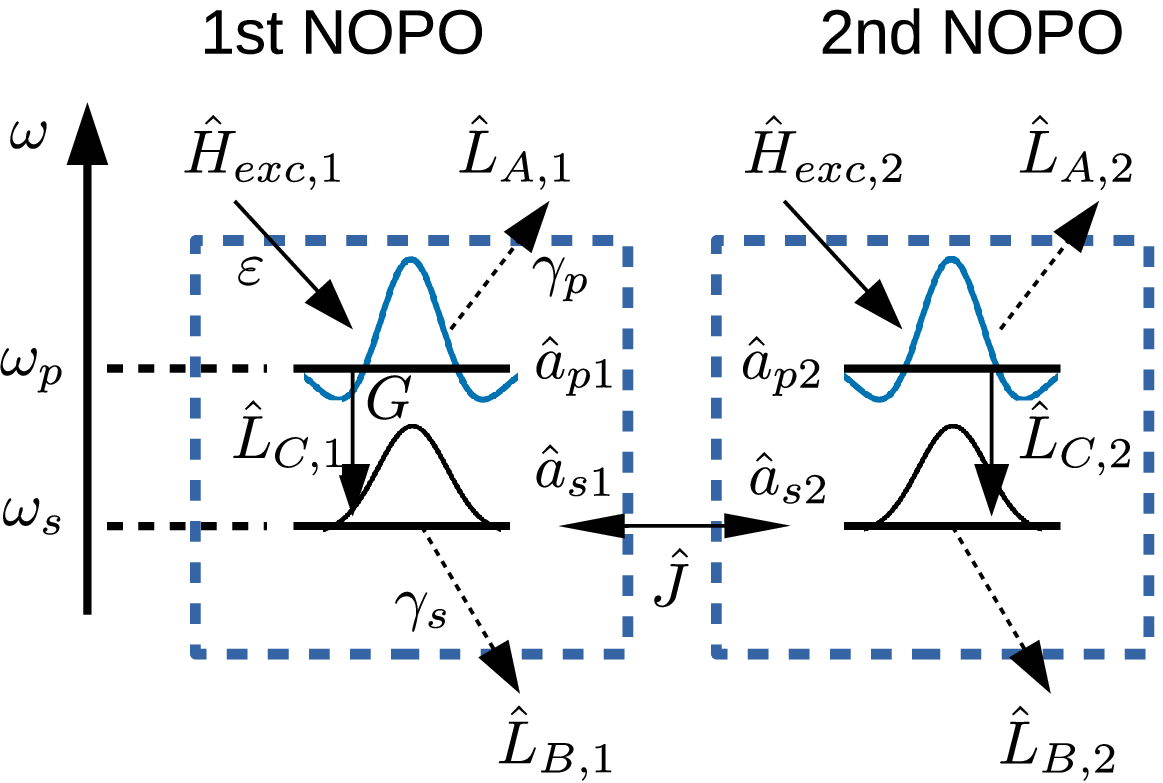}
\caption{Model of dissipatively coupled two NOPOs.}
\label{model}
\end{center}
\end{figure}

We then consider the density matrix master equation of 
dissipatively coupled NOPOs\cite{Takata15} (Fig.\ref{model}). 
The pump mode and signal mode of the $r(=1,2)$-th NOPO are represented by $\hat{a}_{pr}$ and $\hat{a}_{sr}$. 
We assumed that two NOPOs have the identical coefficients $\gamma_p$, $\gamma_s$, $G$ and $\varepsilon$. 
We introduce $\hat{H}_{exc,r}$, $\hat{L}_{A,r}$, $\hat{L}_{B,r}$, and $\hat{L}_{C,r}$ 
as the Hamiltonian and the Liouvillians operating on the modes of $r$-th NOPO. 
The dynamics of the dissipatively coupled two NOPOs is represented by the following master equation: 
\begin{equation}
\label{qmefull}
\frac{\partial \hat{\rho}}{\partial t}=\sum_{r=1,2} \mathcal{L}_{NOPO}^{(r)}\hat{\rho}+([\hat{J},\hat{\rho}\hat{J}^{\dagger}]+{\rm h.c.}). 
\end{equation}
The first term on the right-hand side represents the dynamics of the individual NOPOs, 
and the second term represents the mutual coupling between them. 
The reservoir modes for dissipative interaction, experimentally realized by optical delaylines\cite{Marandi14}, 
are eliminated by the Lindblad procedure\cite{Takata15}. 
Assuming ferromagnetic coupling, we describe $\hat{J}$ in the Liouvillian as follows: 
\begin{equation}
\hat{J}=\sqrt{J}(\hat{a}_{s1}-\hat{a}_{s2}). 
\end{equation}

We next discuss the lasing threshold. 
For a single NOPO, the lasing occurs when the pump photon number becomes $\gamma_s/G$. 
Such pump photons are stored in the pump mode, with the coherent excitation $\varepsilon_{thr}:=\gamma_p\sqrt{\frac{\gamma_s}{G}}$. 
We can use $p=\varepsilon/\varepsilon_{thr}$ for the normalized pump rate. 
If it is smaller than $1$, spontaneous emission is dominant; if 
it is larger than $1$, stimulated emission is dominant. 
At the threshold $p=\varepsilon/\varepsilon_{thr}=1$, 
the extrapolated purely spontaneous emission characteristics reach $\langle\hat{a}_s^{\dagger}\hat{a}_s\rangle=1$. 
Ferromagnetically coupled systems comprising two NOPOs have the same lasing threshold $\varepsilon_{thr}$ as a single NOPO. 
When $G$ is large, the exact threshold differs from the above value. 
In such a case, the required pump photon number for lasing is the same ($\gamma_s/G$). 
It becomes difficult to excite $\gamma_s/G$ photons in the pump mode, however, 
because the spontaneous emission of the pump mode into the signal mode becomes non-negligible. 

\section{Photon-number squeezing in a solitary NOPO}

Here, we present numerical simulation results of photon-number squeezing in a single NOPO. 
Before introducing the numerical methods, we present the analytical result  
in the limit of large dissipation for the pump mode ($\gamma_p\gg \gamma_s$). 
This analytical model shows photon number squeezing. 
For numerical simulation, we use the PSDE \cite{Drummond80b,Gilchrist97}. 
First, we consider the amplitudes of both the pump and signal modes explicitly. 
Such simulation has rigorous correspondence to the original density matrix master equation. 
Then, we derive truncated models with the amplitudes of only the signal mode, 
where the pump mode is adiabatically eliminated under the assumption of the large dissipation. 
Two truncated models with the signal mode represented by the PSDE (T-PSDE), 
or by the WSDE (T-WSDE) are introduced. 

\subsection{Analytical Result}

For a single NOPO, we can derive the analytical steady state  
when the dissipation of the pump mode is sufficiently large\cite{Inui}. 
The analytical result is obtained from the expansion of the density matrix master equation, 
with Glauber's coherent states \cite{Glauber63} for the pump mode $|\alpha_p\rangle$ 
and diagonal Fock states for the signal mode $|N_s\rangle$: 
\begin{equation}
\label{hyb1}
\hat{\rho}=\sum_{N_s=0}^{\infty} \int P_{N_s}(\alpha_p)|\alpha_p\rangle \langle \alpha_p|\otimes |N_s\rangle \langle N_s| d^2\alpha_p. 
\end{equation}
From the Liouvillian of the single NOPO[Eq.(\ref{qme1})], all nondiagonal Fock state components of the signal mode are zero 
when they are zero at the initial state. 
Assuming the vacuum state at $t=0$, we consider only diagonal components of the signal mode. 
Applying the above expansion, we can obtain the following equation for the distribution function $P_{N_s}(\alpha_p)$.
\begin{eqnarray}
\label{hybfp}
\frac{\partial P_{N_s}(\alpha_p)}{\partial t}=\Bigl[ \frac{\partial}{\partial \alpha_p}(\gamma_p+G(1+N_s))\alpha_p P_{N_s} \nonumber \\
-\varepsilon \frac{\partial P_{N_s}}{\partial \alpha_p}+{\rm c.c.}\Bigr]+2\gamma_s[(1+N_s)P_{N_s+1}-N_sP_{N_s}] \nonumber \\
+2G|\alpha_p|^2[N_s P_{N_s-1}-(1+N_s)P_{N_s}]
\end{eqnarray}
Here, we shall neglect the term with the negative signal photon number in the right-hand side. 
This equation represents the drift of $\alpha_p$ and the hopping of $N_s$.  
It is important to notice that, fortunately, there are no terms representing simultaneous drift and hopping. 
If $\gamma_p$ is sufficiently large, we can neglect the dynamics of $\alpha_p$ at the steady state. 
The elimination of the pump mode leads to 
\begin{equation}
\label{dmpe}
P_{N_s}(\alpha_p)=\rho_{N_s}\delta^{(2)}\Bigl(\alpha_p - \frac{\varepsilon}{\gamma_p+G(1+N_s)}\Bigr).
\end{equation}
Integrating Eq.(\ref{hybfp}) with $\int d^2 \alpha_p$, we have 
\begin{eqnarray}
\frac{\partial \rho_{N_s}}{\partial t}=2\gamma_s[(1+N_s)\rho_{N_s+1}-N_s\rho_{N_s}] \nonumber \\
+2[G_e(N_s-1) N_s \rho_{N_s-1}-(1+N_s)G_e(N_s)\rho_{N_s}]. 
\end{eqnarray}
Here, $G_e(N)=G\frac{\varepsilon^2}{[\gamma_p+G(1+N)]^2}$. 
Using the detailed balance in the signal Fock space, we obtain 
\begin{equation}
\frac{\rho_{N_s}}{\rho_{N_s-1}}=\frac{G}{\gamma_s}\frac{\varepsilon^2}{(\gamma_p+GN_s)^2}.
\end{equation}
We note that, in the well-known Scully-Lamb theory\cite{Scully66}, the denominator is proportional to the photon number. 
From the recursion relation, we can obtain 
\begin{equation}
\label{cfan}
\langle \hat{a}_s^{\dagger j}\hat{a}_s^j\rangle=x^j\Gamma(j+1)\frac{\Gamma(c)^2}{\Gamma(j+c)^2}\frac{_1F_2(j+1;j+c,j+c;x)}{_1F_2(1;c,c;x)}
\end{equation}
where $x=\frac{\varepsilon^2}{G\gamma_s}$, $c=1+\frac{\gamma_p}{G}$, $\Gamma(z)$ is the Gamma function, and $_1F_2(\alpha;\beta,\gamma;z)$ is the generalized hypergeometric function. 
From $j=1,2$ of the above expression, we can obtain the mean signal photon number and 
second-order correlation function of the signal mode. 
Far below the threshold ($x\rightarrow 0$), the second order correlation function is 
$g_s^{(2)}(0)=\frac{\langle \hat{a}_s^{\dagger 2}\hat{a}_s^2\rangle}{\langle \hat{a}_s^{\dagger}\hat{a}_s\rangle^2}=2\frac{(\gamma_p+G)^2}{(\gamma_p+2G)^2}$. 
In the $G\rightarrow 0$ limit, this represents $g^{(2)}(0)\rightarrow 2$, which is known as a blackbody radiation state\cite{Walls07}. 

\subsection{Positive-$P$ Representation and PSDE}

The positive-$P$ distribution function of the bosonic mode $\hat{a}$ is defined as follows\cite{Corney03}. 
\begin{equation}
\label{ppex1}
\hat{\rho}=\int P(\alpha,\alpha^{\dagger}) \hat{\Lambda}_P(\alpha,\alpha^{\dagger})d^2\alpha d^2\alpha^{\dagger}, 
\end{equation} 
\begin{equation}
 \hat{\Lambda}_P(\alpha,\alpha^{\dagger})=:e^{-(\hat{a}^{\dagger}-\alpha^{\dagger})(\hat{a}-\alpha)}:
\end{equation}
where $\alpha$ and $\alpha^{\dagger}$ are independent complex amplitudes, 
but their averaged values satisfy $\langle \alpha\rangle^*=\langle \alpha^{\dagger}\rangle$. 
For the master equation [Eq.(\ref{qme1})], we use the expansion. 
\begin{equation}
\hat{\rho}=\int P(\alpha_p,\alpha_p^{\dagger},\alpha_s,\alpha_s^{\dagger}) \hat{\Lambda}_P(\alpha_p,\alpha_p^{\dagger})\otimes \hat{\Lambda}_P(\alpha_s,\alpha_s^{\dagger})dV, 
\end{equation}
where $dV=d^2\alpha_p d^2\alpha_p^{\dagger}d^2\alpha_s d^2\alpha_s^{\dagger}$. 
We obtain the following Fokker-Planck equation of the positive-$P$ distribution function 
$P(\alpha_p,\alpha_p^{\dagger},\alpha_s,\alpha_s^{\dagger})$: 
\begin{eqnarray}
\label{fp2m}
\frac{\partial P}{\partial t}&=&\Bigl[ \frac{\partial}{\partial \alpha_p}(-\varepsilon+\gamma_p \alpha_p+G(1+\alpha_s^{\dagger}\alpha_s)\alpha_p )P \nonumber \\
&+&\frac{\partial}{\partial \alpha_s}(\gamma_s \alpha_s-G\alpha_p^{\dagger}\alpha_p \alpha_s)P-\frac{\partial^2}{\partial \alpha_p\partial \alpha_s}(G\alpha_p\alpha_s P)\nonumber \\
&+&{\rm h.c.}\Bigr]+2G\frac{\partial^2}{\partial \alpha_s^{\dagger}\partial \alpha_s}(\alpha_p^{\dagger}\alpha_p P)
\end{eqnarray}
where ${\rm h.c.}$ represents terms with the operation that transforms the binary 
$[\alpha,\alpha^{\dagger}]^T$ into $[\alpha^{\dagger},\alpha]^T$. 
As the Fokker-Planck equation has derivatives up to the second order, 
the equivalent positive-$P$ stochastic differential equations exist. 
These are obtained via the Ito rule and represented as follows\cite{Inui}. 
\begin{equation}
\label{pp1}
\frac{d \alpha_p}{dt}=-\gamma_p \alpha_p+\varepsilon -G(1+\alpha_s^{\dagger} \alpha_s) \alpha_p -\sqrt{\frac{G}{2}} \alpha_s \xi_{C}^*
\end{equation}
\begin{equation}
\label{pp2}
\frac{d \alpha_s}{dt}=-\gamma_s \alpha_s+G \alpha_p^{\dagger} \alpha_p \alpha_s+ \sqrt{\frac{G}{2}} \alpha_p ( \xi_{C}+\xi_{C}^{\dagger *})
\end{equation}
\begin{equation}
\label{pp3}
\frac{d \alpha_p^{\dagger}}{dt}=-\gamma_p \alpha_p^{\dagger}+\varepsilon -G(1+\alpha_s^{\dagger} \alpha_s) \alpha_p^{\dagger} -\sqrt{\frac{G}{2}} \alpha_s^{\dagger} \xi_{C}^{\dagger *}
\end{equation}
\begin{equation}
\label{pp4}
\frac{d \alpha_s^{\dagger}}{dt}=-\gamma_s \alpha_s^{\dagger}+G \alpha_p^{\dagger} \alpha_p \alpha_s^{\dagger}+ \sqrt{\frac{G}{2}} \alpha_p^{\dagger} ( \xi_{C}^{\dagger}+\xi_{C}^{*})
\end{equation}
The PSDEs for $\alpha_p^{\dagger}$ and $\alpha_s^{\dagger}$ 
are obtained by the simple Hermitian conjugates of those for $\alpha_p$ and $\alpha_s$. 
$\xi_{C}$ represents the complex Gaussian noise with zero means. 
The complex number noise $\xi_{C}$ has the independent Hermitian conjugate $\xi_{C}^{\dagger}$. 
These noise sources have the two time correlation functions $\langle \xi_{C}^*(t)\xi_{C}(t')\rangle=2\delta(t-t')$ and 
$\langle \xi_{C}^{\dagger *}(t)\xi_{C}^{\dagger}(t')\rangle=2\delta(t-t')$. 
The relation between Eq.(\ref{pp1})-(\ref{pp4}) and the PSDE of an NOPO with an explicit idler mode\cite{McNeil83} is shown in Appendix A. 

\subsection{Truncated-PSDE}

The PSDEs in the former section are fully equivalent to the quantum master equation, 
where the mechanism for photon-number squeezing remains unclear, however. 
Here, we take the large $\gamma_p$ limit in Eq.(\ref{pp1}) and Eq.(\ref{pp3}) and eliminate the pump mode adiabatically. 
Such a procedure adds products of random numbers in the PSDE. 
We truncated these terms assuming small $G$. 
\begin{equation}
\label{pae1}
\alpha_p=\frac{\varepsilon}{\Gamma_p}-\frac{1}{\Gamma_p}\sqrt{\frac{G}{2}} \alpha_s \xi^*_{C}
\end{equation}
\begin{equation}
\label{pae2}
\alpha_p^{\dagger}=\frac{\varepsilon}{\Gamma_p}-\frac{1}{\Gamma_p}\sqrt{\frac{G}{2}} \alpha_s^{\dagger} \xi^{\dagger *}_{C}. 
\end{equation}
Here, $\Gamma_p :=\gamma_p+G(1+\alpha_s^{\dagger}\alpha_s)\sim \gamma_p+G\alpha_s^{\dagger}\alpha_s$. 
We substitute Eqs.(\ref{pae1})(\ref{pae2}) into Eqs.(\ref{pp2})(\ref{pp4}) and ignore the products of noise sources. 
After eliminating $\alpha_p$ and $\alpha_p^{\dagger}$, we can obtain the following Fokker-Planck equation of 
$P(\alpha_s,\alpha_s^{\dagger})=\int P(\alpha_p,\alpha_p^{\dagger},\alpha_s,\alpha_s^{\dagger})d^2\alpha_p d^2\alpha_p^{\dagger}$ 
which is equivalent to the stochastic differential equations of $\alpha_s$ and $\alpha_s^{\dagger}$. 
\begin{eqnarray}
\label{fppp}
\frac{\partial P}{\partial t}&=&\Bigl[ \frac{\partial}{\partial \alpha_s}(\gamma_s \alpha_s -G_e\alpha_s)P-\frac{\partial^2}{\partial \alpha_s^2}\Bigl(\frac{F_e}{2}\alpha_s^2 P\Bigr)+{\rm h.c.}\Bigr]\nonumber \\
&+&2\frac{\partial^2}{\partial \alpha_s^{\dagger}\partial \alpha_s}\Bigl(G_e-\frac{F_e}{2}\alpha_s^{\dagger}\alpha_s\Bigr)P
\end{eqnarray}
Here, $G_e=\frac{\varepsilon^2G}{\Gamma_p^2}$, and $F_e=\frac{2\varepsilon^2 G^2}{\Gamma_p^3}$. 
$G_e$ represents the gain coefficient which contains the effect of pump depletion. 
The terms with $G_e$ resemble the spontaneous and stimulated emission terms in Ref.\cite{Walls73}. 
On the other hand, the terms with $F_e$, which are $ -(\frac{\partial^2}{\partial \alpha^2}(\alpha^2 P)+{\rm c.c.})
-2\frac{\partial}{\partial \alpha \partial \alpha^*}(|\alpha|^2 P) $ in Glauber's representation, 
do not correspond to a simple Liouvillian 
(Following Lax-Louisell's procedure\cite{Lax69}, they correspond to the complicated Liouvillian 
$ \mathcal{L}\hat{\rho}\sim (\hat{a}^{\dagger 2}\hat{a}^2\hat{\rho}-4\hat{a}^{\dagger}\hat{a}^2\hat{\rho}\hat{a}^{\dagger}+{\rm h.c.})+4\hat{a}^2\hat{\rho}\hat{a}^{\dagger 2}+2\hat{a}^{\dagger}\hat{a}\hat{\rho}\hat{a}^{\dagger}\hat{a} -2\hat{a}\hat{\rho}\hat{a}^{\dagger}$.).
We can point out that the resulting Fokker-Planck equation resembles that of the simple bosonic dephasing Liouvillian\cite{Agarwal78}: 
$\mathcal{L}_{ph}\hat{\rho}\sim [\hat{a}^{\dagger}\hat{a},\hat{\rho}\hat{a}^{\dagger}\hat{a}]+{\rm h.c.}$, 
which has the Fokker-Planck equation 
$\sim \Bigl[\frac{\partial (\alpha P)}{\partial \alpha}-\frac{\partial^2}{\partial \alpha^2}(\alpha^2 P)+{\rm c.c.}\Bigr]+2\frac{\partial^2 }{\partial \alpha^{*} \partial \alpha}(|\alpha|^2 P)$. 
In the polar representation of the coherent state $\alpha=\sqrt{I}e^{i\theta}$, with 
\begin{equation}
\label{polar}
\frac{\partial}{\partial \alpha}\alpha=\frac{\partial}{\partial I}I-\frac{i}{2}\frac{\partial}{\partial \theta}, 
\end{equation}
the dephasing process contains the squared imaginary part 
$(\frac{\partial}{\partial \alpha}\alpha-\frac{\partial}{\partial\alpha^{*}}\alpha^{*})^2P$. 
On the other hand, the photon number fluctuation contains the squared real part 
$(\frac{\partial}{\partial \alpha}\alpha+\frac{\partial}{\partial\alpha^{*}}\alpha^{*})^2P$. 
Since the terms with $F_e$ in Eq.(\ref{fppp}) have the negative-sign squared real part of Eq.(\ref{polar}), 
they represent the removal of photon number fluctuation, which yields the photon-number-squeezed state. 
The positive-$P$ stochastic differential equations for only the signal mode are obtained as follows:
\begin{equation}
\label{pp5}
\frac{d\alpha_s}{dt}=-\gamma_s\alpha_s+G_e\alpha_s+\sqrt{G_f}\xi_{C}+i\sqrt{F_e}\alpha_s\xi_{R}
\end{equation}
\begin{equation}
\label{pp6}
\frac{d\alpha_s^{\dagger}}{dt}=-\gamma_s\alpha_s^{\dagger}+G_e\alpha_s^{\dagger}+\sqrt{G_f}\xi_{C}^*-i\sqrt{F_e}\alpha_s^{\dagger}\xi_{R}^{\dagger}
\end{equation}
Here, $G_f=G_e-\frac{F_e}{2}\alpha_s^{\dagger}\alpha_s\sim \frac{\varepsilon^2 G\gamma_p}{\Gamma_p^3}$. 
The complex number noise source $\xi_C$ satisfies $\langle \xi_C^*(t)\xi_C(t')\rangle=2\delta(t-t')$. 
$\xi_{R}$ and $\xi_{R}^{\dagger}$ are independent real-number noise sources with 
$\langle \xi_{R}(t)\xi_{R}(t')\rangle=\delta(t-t')$ and $\langle \xi^{\dagger}_{R}(t)\xi_{R}^{\dagger}(t')\rangle=\delta(t-t')$.  
These truncated PSDEs for only signal mode represent $g_s^{(2)}(0)=2$ for spontaneous emission where $G_e=G_f$ and $\sqrt{F_e}\alpha_s$ is small. 
Due to the truncation process, they cannot represent analytical $g_s^{(2)}(0)=2\frac{(\gamma_p+G)^2}{(\gamma_p+2G)^2}$, which deviates from $2$ when $G$ is large. 

\begin{figure*}
\begin{center}
\includegraphics[width=16.0cm]{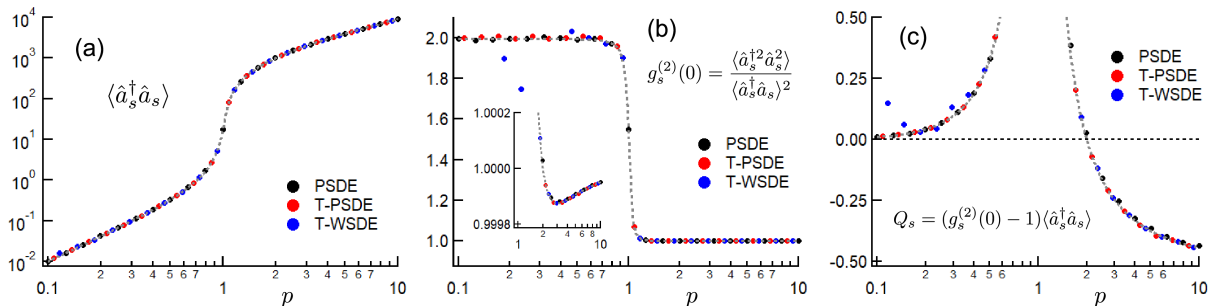}
\caption{Numerical results for steady-state single NOPO with $\gamma_p \gg \gamma_s$ 
as a function of normalized excitation $p=\varepsilon/\varepsilon_{thr}$. 
(a) Mean signal photon number $\langle \hat{a}_s^{\dagger}\hat{a}_s\rangle$. 
(b) Second order correlation function of the signal mode: $g_s^{(2)}(0)$. 
(c) Mandel's $Q$ parameter of the signal mode. 
Gray broken lines are analytical results. }
\label{ss1}
\end{center}
\end{figure*}

We numerically integrate the PSDE with the explicit pump mode 
[Eqs.(\ref{pp1})(\ref{pp2})(\ref{pp3})(\ref{pp4})], 
and the truncated-PSDE without the explicit pump mode [Eqs.(\ref{pp5})(\ref{pp6})]. 
We compare the results obtained by these two models with the analytical results [Eq.(\ref{cfan})]. 
The numerical results obtained by PSDEs are shown in Fig.\ref{ss1}, together with analytical results. 
The numerical results were calculated with $9\times 10^4$ trajectories. 
Each trajectory starts from a vacuum state. 
For the first $10^2/\gamma_s$ time period, the excitation was adiabatically increased as $p(t)=p \sqrt{\gamma_s t/10^2}$, 
where $p$ is the excitation at the steady state. 
After that time, the time average was taken for another $10^2/\gamma_s$ time period where $p$ is constant. 
The results are averaged over the $10^2/\gamma_s$ time period and $9\times 10^4$ trajectories. 
The gray broken line in Fig. \ref{ss1}(a) represents the analytical mean signal photon number. 
The signal photon number increases nonlinearly at the threshold $p=\varepsilon/\varepsilon_{thr}=1$. 
The gray broken line in Fig.\ref{ss1}(b) represents the analytical second-order correlation function of the signal mode 
$g_s^{(2)}(0)=\frac{\langle \hat{a}_s^{\dagger 2}\hat{a}_s^2\rangle}{\langle \hat{a}_s^{\dagger}\hat{a}_s\rangle^2}$ obtained by Eq.(\ref{cfan}). 
Below the threshold, this value is almost equal to $2$, which shows that the signal mode is in the blackbody radiation state\cite{Walls07}.  
The signal mode turns from the blackbody radiation state below the threshold into the coherent state above the threshold. 
At the threshold, $g_s^{(2)}(0)\sim \frac{\pi}{2}$\cite{Risken67} is obtained 
due to the non-Gaussian Glauber distribution function $\log P(\alpha_s)\sim -|\alpha_s|^4$. 
Above the threshold, $g^{(2)}_s (0)$ is almost $1$ but, as shown in the inset, 
is slightly smaller than $1$ due to the photon-number squeezing. 
Below the threshold, the values of $g^{(2)}(0)$ are slightly smaller than $2$, 
for analytical results and PSDE with the explicit pump mode. 
These methods can consider the correction of small-$G$, which was neglected in the T-PSDE. 

To evaluate the photon-number squeezing occurring above the threshold, 
we introduce Mandel's $Q$ parameter of the signal mode\cite{Mandel79}: 
\begin{equation}
Q_s=\frac{\langle (\Delta \hat{a}_s^{\dagger}\hat{a}_s)^2 \rangle-\langle \hat{a}_s^{\dagger}\hat{a}_s\rangle}{\langle \hat{a}_s^{\dagger}\hat{a}_s\rangle}.
\end{equation}
Here $\Delta \hat{a}_s^{\dagger}\hat{a}_s=\hat{a}_s^{\dagger}\hat{a}_s-\langle \hat{a}_s^{\dagger}\hat{a}_s\rangle$. 
The coherent state has $Q_s=0$, and Fock state has $Q_s=-1$. 
The photon-number-squeezed state has $Q_s$ smaller than $0$. 
Fig.\ref{ss1}(c) shows that at the threshold $Q_s$ has the maximal value and 
that above the threshold $Q_s$ decreases monotonically to below $Q_s=0$. 
We can see that a single NOPO in the large pump linewidth limit produces a photon-number-squeezed state 
with the pump rates $\varepsilon/\varepsilon_{thr}>2$. 
From the analytical results [Eq.(\ref{cfan})], $Q_s$ converges to $Q\rightarrow -0.5$, as predicted far above the threshold 
for a model with electronic levels\cite{Ritsch92,Gheri92}. 
The results of the PSDE with the explicit pump mode have slightly larger $Q_s$ above the threshold. 
This is because the rigorous PSDE calculation contains the correction of $Q_s$ due to the finite $\gamma_p/\gamma_s$. 

\subsection{Wigner Representation and Truncated-WSDE}

We also consider the truncated Wigner representation for the density operator. 
In the final form, this approach can provide the SDE of only one complex variable for each OPO. 
However, at first, instead of the ordinary Wigner function, 
we use the positive-Wigner function defined by the following expansion of the density matrix \cite{Corney03}. 
\begin{equation}
\hat{\rho}=\int W(\alpha,\alpha^{\dagger}) \hat{\Lambda}_W(\alpha,\alpha^{\dagger})d^2\alpha d^2\alpha^{\dagger},
\end{equation}
where
\begin{equation}
 \hat{\Lambda}_W(\alpha,\alpha^{\dagger})=2:e^{-2(\hat{a}^{\dagger}-\alpha^{\dagger})(\hat{a}-\alpha)}: .
\end{equation}
We expand the density operator using the positive-Wigner function for the signal mode, 
while the pump mode is expanded with positive-$P$ distribution function:  
\begin{equation}
\hat{\rho}=\int \Omega(\alpha_p,\alpha_p^{\dagger},\alpha_s,\alpha_s^{\dagger})\hat{\Lambda}_P(\alpha_p,\alpha_p^{\dagger})\otimes \hat{\Lambda}_W(\alpha_s,\alpha_s^{\dagger})dV.
\end{equation} 
Here, $dV=d^2\alpha_p d^2\alpha_p^{\dagger}d^2\alpha_s d^2\alpha_s^{\dagger}$, and 
$\Omega$ is the hybrid distribution function, which produces a positive Wigner function when the pump mode is integrated:
$W(\alpha_s,\alpha_s^{\dagger})=\int \Omega(\alpha_p,\alpha_p^{\dagger},\alpha_s,\alpha_s^{\dagger})d^2\alpha_p d^2\alpha_p^{\dagger}$.
The Fokker-Planck equation under the above expansion has third-order derivatives. 
We neglect these derivatives, and obtain the following stochastic differential equations for the pump amplitude and signal amplitude. 
\begin{equation}
\frac{d\alpha_p}{dt}=-\gamma_p\alpha_p+\varepsilon-G\Bigl(\alpha_s^{\dagger}\alpha_s+\frac{1}{2}\Bigr)\alpha_p-\sqrt{\frac{G}{4}}(\alpha_s \xi_{C1}+\alpha_s^{\dagger}\xi_{C2})
\end{equation}
\begin{eqnarray}
\frac{d\alpha_s}{dt}=-\gamma_s\alpha_s+G\alpha_p^{\dagger}\alpha_p\alpha_s+\sqrt{\frac{\gamma_s}{2}+\frac{G}{2}\alpha_p^{\dagger}\alpha_p}\xi_C\nonumber \\
+\sqrt{\frac{G}{4}}(\alpha_p \xi_{C1}^*-\alpha_p^{\dagger}\xi_{C2}^{\dagger *})
\end{eqnarray}
\begin{equation}
\frac{d\alpha_p^{\dagger}}{dt}=-\gamma_p\alpha_p^{\dagger}+\varepsilon-G\Bigl(\alpha_s^{\dagger}\alpha_s+\frac{1}{2}\Bigr)\alpha_p^{\dagger}-\sqrt{\frac{G}{4}}(\alpha_s^{\dagger} \xi_{C1}^{\dagger}+\alpha_s\xi_{C2}^{\dagger})
\end{equation}
\begin{eqnarray}
\frac{d\alpha_s^{\dagger}}{dt}=-\gamma_s\alpha_s^{\dagger}+G\alpha_p^{\dagger}\alpha_p\alpha_s^{\dagger}+\sqrt{\frac{\gamma_s}{2}+\frac{G}{2}\alpha_p^{\dagger}\alpha_p}\xi_C^*\nonumber \\
+\sqrt{\frac{G}{4}}(\alpha_p^{\dagger} \xi_{C1}^{\dagger *}-\alpha_p\xi_{C2}^{*})
\end{eqnarray}
$\xi_C$, $\xi_{C1}$, $\xi_{C1}^{\dagger}$, $\xi_{C2}$ and $\xi_{C2}^{\dagger}$ are independent complex random numbers. 
We eliminate the pump mode in a way similar to Eqs.(\ref{pae1}) and (\ref{pae2}) 
and ignore the product of noise terms. 
After the elimination of the pump mode, we do not require the positive-Wigner representation. 
We obtain the following stochastic differential equation in Wigner representation: 
\begin{equation}
\label{wig1}
\frac{d\alpha_s}{dt}=-\gamma_s\alpha_s+G_e\alpha_s+\sqrt{\frac{\gamma_s}{2}+\frac{G_e}{2}}\xi_C
\end{equation}
where $\xi_C$ is a complex random number with $\langle \xi_C^*(t)\xi_C(t')\rangle=2\delta(t-t')$. 
Here, $G_e$ is the same as that in the T-PSDE with $\Gamma_p=\gamma_p+G|\alpha_s|^2$ 
under the assumption of $\gamma_p \gg G$. 
The $\langle \hat{a}_s^{\dagger}\hat{a}_s\rangle$, $g^{(2)}_s(0)$ and $Q_s$ 
obtained with the T-WSDE are represented in Fig.\ref{ss1}. 
Results for the T-WSDE above the threshold are similar to those of the T-PSDE in Fig.\ref{ss1}, 
although far below the threshold, the results with the truncated WSDE have large fluctuation. 
We used Eq.(\ref{wig1}) to derive analytical results in Appendix B. 

\section{Entanglement in two coupled NOPOs}

\subsection{Small-gain Case}

\begin{figure}[h]
\begin{center}
\includegraphics[width=8.0cm]{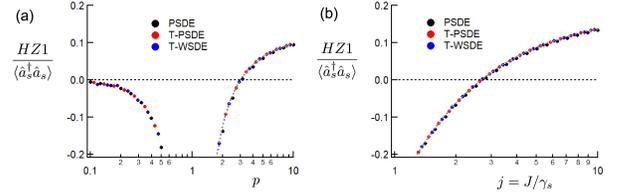}
\caption{Normalized $HZ1$ value in the dissipatively coupled two NOPOs 
(a) as a function $p=\varepsilon/\varepsilon_{thr}$ with constant coupling coefficient $j=4$, and 
(b) as a function $j=J/\gamma_s$ with constant excitation $p=5$. 
Gray broken lines are analytical results shown in Appendix B. } 
\label{ds1}
\end{center}
\end{figure}

We now consider the system of two NOPOs with dissipative coupling coefficient $J$, 
below normalized as $j=J/\gamma_s$. 
We consider the network of two NOPOs where two signal modes represented by $\hat{a}_{s1}$ and $\hat{a}_{s2}$ are dissipatively coupled. 
To quantify the entanglement between two signal modes, we use one of the Hillery-Zubairy criteria\cite{Hillery06} which we call $HZ1$. 
\begin{equation}
HZ1=|\langle \hat{a}^{\dagger}_{s1}\hat{a}_{s2}\rangle|^2 -\langle \hat{a}_{s1}^{\dagger}\hat{a}_{s1}\hat{a}_{s2}^{\dagger}\hat{a}_{s2}\rangle
\end{equation}
$HZ1>0$ is the sufficient condition for entanglement. 
If the density matrix after tracing out the pump modes 
can be separated into that of the first NOPO ($\hat{\rho}_1$) and that of the second NOPO ($\hat{\rho}_2$), 
as $\hat{\rho}=\hat{\rho}_{1}\otimes \hat{\rho}_{2}$, 
$HZ1 \le 0$ can be obtained from the Schwarz inequality. 
Therefore, when $HZ1>0$ the system of two NOPOs is not in the product state. 
The $HZ1$ value can be calculated with PSDE, T-PSDE and T-WSDE. 
Calculation was performed for two NOPOs with $G/\gamma_s=0.05$, $\gamma_p/\gamma_s=50$ and $j=4$. 
The steady state results were obtained in the same way as those in Fig.\ref{ss1}. 
We plot the $HZ1$ value normalized by mean photon number $\langle \hat{a}_s^{\dagger}\hat{a}_s\rangle$ 
as a function of the normalized pump rate $p=\varepsilon/\varepsilon_{thr}$ in Fig.\ref{ds1}(a). 
$HZ1$ is negative just above the threshold but becomes positive at $p\sim 3$.    
We can see that the dissipatively coupled NOPOs can satisfy the entanglement criterion. 
The gray broken line represents the analytical results shown in Appendix B. 
The $j$ dependence of $HZ1$ is shown in Fig.\ref{ds1}(b) with the excitation $p=5$. 
When $j$ is small, the entanglement criterion is not fulfilled, 
although the photon-number-squeezed state is attained in each NOPO. 
$HZ1$ becomes positive only for large $j$. 
In Fig.\ref{ds1}(b), $j \sim 8/3$ is required. 
The PSDE with the explicit pump mode had a slightly larger $Q_s$ in Fig.\ref{ss1}(c), and also a slightly smaller $HZ1$ value.  

\subsection{Large-gain Case and Non-diagonal Components}

\begin{figure*}
\begin{center}
\includegraphics[width=14.0cm]{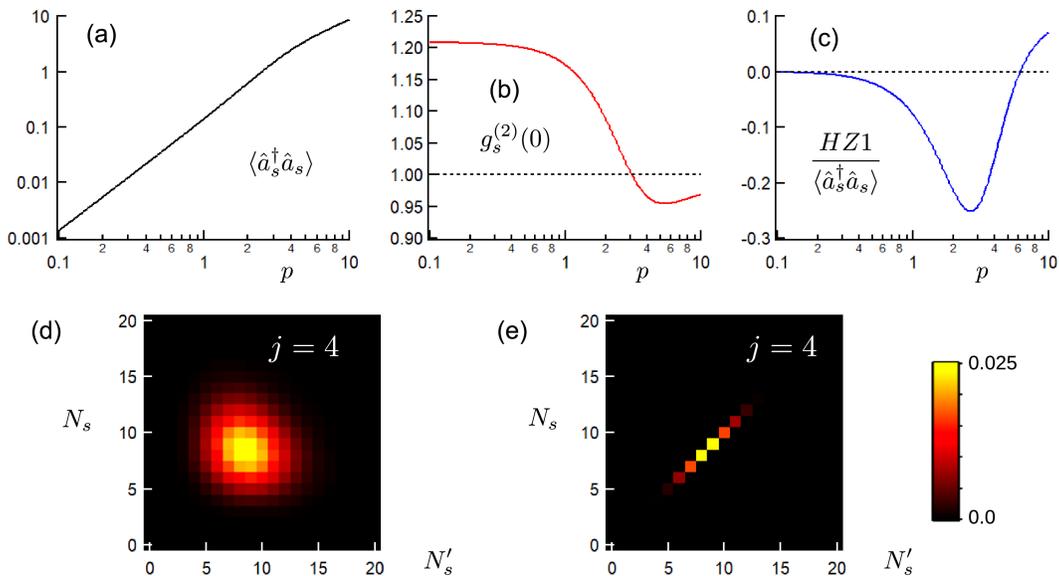}
\caption{Steady state of two coupled NOPOs with large gain coefficient $G$. 
Excitation dependent (a) mean signal photon number $\langle \hat{a}_s^{\dagger}\hat{a}_s\rangle$, 
(b) second-order correlation function $g_s^{(2)}(0)$,  
and (c) normalized $HZ1$ value. The dissipative coupling coefficient was $j=4$. 
Nondiagonal Fock space components (d) $\langle N_s,N_s'|\hat{\rho}|N_s',N_s \rangle$ 
and (e) $\langle N_s,N_s|\hat{\rho}|N_s',N_s'\rangle$ with $p=10$ and $j=4$. }
\label{nd1}
\end{center}
\end{figure*}

Here we describe the entanglement formed in the two coupled NOPOs 
by calculating non-diagonal elements in a Fock state representation. 
The density matrix equation for a single NOPO is Eq.(\ref{qme1}). 
We perform the direct calculation of the density matrix equation 
after eliminating the pump mode in a way similar to Eq.(\ref{dmpe}). 
For a single NOPO, the density matrix components for the signal photon number $N_s$ 
($\hat{\rho}=\sum_{N_s,N_s'}\rho_{N_s,N_s'}|N_s\rangle \langle N_s'|$) have the following equation: 
\begin{widetext}
\begin{eqnarray}
\label{fock2m}
\frac{\partial \rho_{N_s,N_s'}}{\partial t}&=&2\gamma_s\sqrt{(N_s+1)(N_s'+1)}\rho_{N_s+1,N_s'+1}-\gamma_s(N_s+N_s')\rho_{N_s,N_s'} \nonumber \\
&+&2G_e(N_s-1,N_s'-1)\sqrt{N_s N_s'}\rho_{N_s-1,N_s'-1}-[G_e(N_s,N_s')(N_s+1)+G_e(N_s,N_s')(N_s'+1)]\rho_{N_s,N_s'}.
\end{eqnarray}
\end{widetext}
Here, $G_e(M,N)=G\frac{\varepsilon^2}{[\gamma_p+G(1+M)][\gamma_p+G(1+N)]}$, 
and we shall neglect terms with a negative photon number in the right-hand side. 
We can easily extend this equation to the dissipatively coupled two NOPOs. 
The derivation and validation of this numerical method is shown in Appendix C. 
To obtain the photon-number-squeezed state with the small photon number 
available in density matrix calculation, we take a large $G$ value. 
We calculate the steady-state of the two coupled NOPOs with $G/\gamma_s=50$ and $\gamma_p/\gamma_s=50$. 
The time development started from the vacuum state. 
The excitation depended on time as $p\min(1,\sqrt{t\gamma_s/10})$. 
The state after the $40/\gamma_s$ time period was evaluated. 
The maximum photon number considered was $N_s=30$ and states with photon number larger than that were neglected. 
First, the excitation dependence was studied with a fixed dissipative coupling constant $j=4$. 
The excitation-dependent mean signal photon number 
$\langle \hat{a}_s^{\dagger}\hat{a}_s\rangle$ is shown in Fig.\ref{nd1}(a). 
We can see that the signal photon number does not show a nonlinear increase at $\langle \hat{a}_s^{\dagger}\hat{a}_s\rangle=1$. 
This thresholdless behavior is due to the condition of the unity saturation parameter\cite{Bjork94}. 
We present the second-order correlation function $g_s^{(2)}(0)$ in Fig.\ref{nd1}(b). 
Below the threshold, the single NOPO has $g_s^{(2)}(0)=8/9$ for $G=\gamma_p$. 
However, we obtained the larger value $g_s^{(2)}(0)\sim 1.21$ due to the dissipative coupling. 
We can see that $g_s^{(2)}(0)$ changes explicitly from $g_s^{(2)}(0)>1$ below the threshold to $g_s^{(2)}(0)<1$ above the threshold. 
Next, the $HZ1$ entanglement criterion was calculated. 
The positive $HZ1$, which shows the inseparability of the two NOPOs, is shown in Fig.\ref{nd1}(c).  
We can see that for large $G$, the entanglement formation requires a larger $p$ value 
than in the small-$G$ case in Fig.\ref{ds1}, 
due to the change in an effective threshold in such a thresholdless oscillator. 

We plot the nondiagonal Fock space components $\langle N_s,N_s'|\hat{\rho}|N_s',N_s\rangle$ in Fig.\ref{nd1}(d) for $p=10$ and $j=4$, 
where $HZ1>0$ and the mean signal photon number was $\langle \hat{a}_s^{\dagger}\hat{a}_s\rangle\sim 8.72$. 
The peak of the plot is larger than that of the coherent state 
$(\frac{\langle \hat{a}_s^{\dagger}\hat{a}_s\rangle^{N_s}}{N_s!}e^{-\langle \hat{a}_s^{\dagger}\hat{a}_s\rangle})^2$ 
due to the photon-number-squeezing factor $1/(1+Q_s)$. Here $Q_s\sim -0.27$. 
With large mutual coupling satisfying $HZ1>0$, as seen in Fig.\ref{nd1}(d), 
the superposition of Fock states occurs sufficiently for states with 
photon number difference $\lesssim \sqrt{\langle \hat{a}_s^{\dagger}\hat{a}_s\rangle}$ around the mean photon number. 
The growth of the non-diagonal components indicates the formation of superposition 
between Fock states with the same total signal photon number. 
We then present the non-diagonal components $\langle N_s,N_s|\hat{\rho}|N_s',N_s'\rangle$. 
Such components contribute to the entanglement between the signal and idler mode in a single NOPO\cite{Walls07}. 
As shown in Fig.\ref{nd1}(e), such non-diagonal components are absent 
in signal modes of dissipatively coupled photon-number-squeezed states. 

\section{Summary}

We numerically showed that the steady-state photon-number squeezing 
is attained in a solitary NOPO above the threshold, 
and that such squeezed states result in quantum entanglement 
with sufficiently large dissipative coupling between two NOPOs. 
We introduced stochastic differential equations, 
and the numerical method with Fock space components, for Shen's Raman laser model.  
When the gain coefficient per pump photon ($G$) is small, the PSDE, truncated PSDE, and the truncated WSDE show 
similar results on photon number squeezing and entanglement. 
The truncated PSDE after the adiabatic elimination of the pump mode has a corresponding Fokker-Planck equation 
that explicitly represents the nonclassical negative photon number diffusion. 
The truncated WSDE provides a simple equation which represents the photon number squeezed state. 
We showed that the two coupled NOPOs pass the Hillery-Zubairy's criterion of entanglement, 
when each NOPO operates far above the threshold and 
the dissipative linear coupling is sufficiently large. 
We considered two coupled NOPOs with large gain coefficient $G$, 
with the numerical calculation of the Fock space components after eliminating the pump modes. 
When the system satisfies the entanglement criterion, 
we can see large non-diagonal Fock space components. 

For a single NOPO with large $G$, antibunching with $g^{(2)} (0)\rightarrow 0.5$ was obtained in the spontaneous emission regime. 
This nonclassicality shows possible entanglement even below the threshold. 
We did not show positive $HZ1$ for NOPOs below the threshold, however, 
even when a solitary NOPO can satisfy $g^{(2)}(0)<1$. 
To obtain $HZ1>0$ below threshold, we note that smaller correlation function 
$g^{(2)}(0)\rightarrow 0$ must be achieved\cite{Pellizzari94} 
when the detuning exists between pump, signal and electronic excitation. 
The detuning could be introduced in our scheme with the nondegenerate Kerr interaction, 
i.e., cross-phase modulation between the pump and signal waves, $\hat{H}\sim \hat{a}_p^{\dagger}\hat{a}_p\hat{a}_s^{\dagger}\hat{a}_s$, 
that is added when pump-signal-idler coupling has the detuning\cite{Holland90}. 

We only consider an NOPO via second-order nonlinear susceptibility $\chi^{(2)}$, 
although experimentally a nonlinear optical realization of a coherent XY machine was 
achieved with hyperparametric oscillation via degenerate four-wave-mixing\cite{Takeda17}. 
The pump mode of the hyperparametric oscillator has, above the threshold, two-photon-absorption into the signal mode. 
As two-photon-absorption of the coherently excited mode realizes 
photon number squeezed state with Mandel's $Q\rightarrow -1/3$ \cite{Chaturvedi77}, 
the photon number squeezing of the signal mode will be enhanced. 

\begin{acknowledgments}

Y.I. thanks K.Kamide, T.Asano, and S.Noda for comments on the solitary Raman laser model. 
This project was supported by the ImPACT program of the Japanese cabinet office. 

\end{acknowledgments}

\appendix

\section{Relation between NOPO model and Shen's model}

Here, we present the relation between the NOPO model and Shen's model with adiabatic elimination of the idler mode. 
The well-known PSDE of NOPO with Eq.(\ref{sNOPO}) is as follows\cite{McNeil83,Reid88}:
\begin{equation}
\label{app_p}
\frac{d\alpha_p}{dt}=-\gamma_p \alpha_p+\varepsilon-\kappa\alpha_s\alpha_i
\end{equation}
\begin{equation}
\frac{d\alpha_s}{dt}=-\gamma_s \alpha_s+\kappa\alpha_i^{\dagger}\alpha_p+\sqrt{\frac{\kappa}{2}\alpha_p}\xi_{C}
\end{equation}
\begin{equation}
\frac{d\alpha_i}{dt}=-\gamma_i \alpha_i+\kappa\alpha_s^{\dagger}\alpha_p+\sqrt{\frac{\kappa}{2}\alpha_p}\xi_{C}^*
\end{equation}
\begin{equation}
\label{app_pT}
\frac{d\alpha_p^{\dagger}}{dt}=-\gamma_p \alpha_p^{\dagger}+\varepsilon-\kappa\alpha_s^{\dagger}\alpha_i^{\dagger}
\end{equation}
\begin{equation}
\frac{d\alpha_s^{\dagger}}{dt}=-\gamma_s \alpha_s^{\dagger}+\kappa\alpha_i\alpha_p^{\dagger}+\sqrt{\frac{\kappa}{2}\alpha_p^{\dagger}}\xi_{C}^{\dagger}
\end{equation}
\begin{equation}
\frac{d\alpha_i^{\dagger}}{dt}=-\gamma_i \alpha_i^{\dagger}+\kappa\alpha_s\alpha_p^{\dagger}+\sqrt{\frac{\kappa}{2}\alpha_p^{\dagger}}\xi_{C}^{\dagger *}
\end{equation}
After $\alpha_i$ and $\alpha_i^{\dagger}$ are adiabatically eliminated as 
$\alpha_i=\frac{\kappa}{\gamma_i}\alpha_s^{\dagger}\alpha_p+\sqrt{\frac{\kappa}{2\gamma_i^2}\alpha_p}\xi_{C}^{*}$ and 
$\alpha_i^{\dagger}=\frac{\kappa}{\gamma_i}\alpha_s\alpha_p^{\dagger}+\sqrt{\frac{\kappa}{2\gamma_i^2}\alpha_p^{\dagger}}\xi_{C}^{\dagger *}$, 
the corresponding Fokker-Planck equation of $P(\alpha_p,\alpha_p^{\dagger},\alpha_s,\alpha_s^{\dagger})$ 
is similar to Eq.(\ref{fp2m}) with $G=\kappa^2/\gamma_i$. 
However, the SDEs do not correspond directly to Eqs.(\ref{pp1})-(\ref{pp4}). 
Here, it is known that we can choose the diffusion coefficients arbitrarily\cite{Gilchrist97}. 
With the pump mode following Eqs.(\ref{app_p})(\ref{app_pT}), we can obtain the following SDEs for signal and idler modes. 
\begin{equation}
\label{app_s}
\frac{d\alpha_s}{dt}=-\gamma_s \alpha_s+\kappa\alpha_i^{\dagger}\alpha_p+\sqrt{\frac{G}{2}}\alpha_p\xi_{C}
\end{equation}
\begin{equation}
\label{app_i}
\frac{d\alpha_i}{dt}=-\gamma_i \alpha_i+\kappa\alpha_s^{\dagger}\alpha_p+\sqrt{\frac{\gamma_i}{2}}\xi_{C}^*
\end{equation}
\begin{equation}
\label{app_sT}
\frac{d\alpha_s^{\dagger}}{dt}=-\gamma_s \alpha_s^{\dagger}+\kappa\alpha_i\alpha_p^{\dagger}+\sqrt{\frac{G}{2}}\alpha_p^{\dagger} \xi_{C}^{\dagger}
\end{equation}
\begin{equation}
\label{app_iT}
\frac{d\alpha_i^{\dagger}}{dt}=-\gamma_i \alpha_i^{\dagger}+\kappa\alpha_s\alpha_p^{\dagger}+\sqrt{\frac{\gamma_i}{2}}\xi_{C}^{\dagger *}
\end{equation}
When the idler mode is adiabatically eliminated with 
$\alpha_i=\frac{\kappa}{\gamma_i}\alpha_s^{\dagger}\alpha_p+\sqrt{\frac{1}{2\gamma_i}}\xi_C^*$ and 
$\alpha_i^{\dagger}=\frac{\kappa}{\gamma_i}\alpha_s\alpha_p^{\dagger}+\sqrt{\frac{1}{2\gamma_i}}\xi_C^{\dagger *}$, 
from Eqs.(\ref{app_p})(\ref{app_s})(\ref{app_pT})(\ref{app_sT}), 
Eqs.(\ref{pp1})-(\ref{pp4}) without the spontaneous emission loss of the pump mode are obtained directly. 
The spontaneous emission loss of the pump mode appears in the time development of 
the amplitude products $\alpha_i \alpha_s$ and $\alpha_i^{\dagger}\alpha_s^{\dagger}$, 
but it is not considered when we eliminate the idler mode from PSDE.  

\section{Analytical result for small-gain case}

In a recent paper on a coherent Ising machine, we showed an analytical method for calculating entanglement criterion\cite{Inui19}. 
Here, we apply this method for dissipatively coupled two NOPOs above the threshold. 
We start the discussion from the T-WSDE, assuming large $\gamma_p$. 
First, we consider a solitary NOPO. 
Above the threshold, in the square root of Eq.(\ref{wig1}), we assume $G_e\sim \gamma_s$. 
We consider the fluctuation around the mean amplitude $\langle \alpha_s\rangle=\sqrt{\frac{\gamma_p}{G}(p-1)}$. 
For the fluctuation part $\Delta \alpha_s=\alpha_s-\langle \alpha_s\rangle$, 
the drift term is $-\gamma_s\Delta \alpha_s+G_e\Delta \alpha_s+\langle \alpha_s\rangle \Delta G_e \sim \langle \alpha_s\rangle \Delta G_e$.
Here, $\Delta G_e\sim -2\frac{G^2\varepsilon^2}{\Gamma_p^3}\langle \alpha_s\rangle(\Delta \alpha_s+\Delta \alpha_s^{*})$.  
The fluctuation part $\Delta \alpha_s$ has the following equation. 
\begin{equation}
\frac{d\Delta \alpha_s}{dt}=-2\gamma_s \Bigl(1-\frac{1}{p}\Bigr) (\Delta \alpha_s+\Delta \alpha_s^{*})+\sqrt{\gamma_s}\xi_s
\end{equation}
Mandel's $Q$ parameter satisfies $Q_s=\langle (\Delta \alpha_s+\Delta \alpha_s^{*})^2\rangle-1$ and 
is written as $Q_s=-\frac{1}{2}+\frac{1}{2(p-1)}$. 
Therefore, when $p>2$ the amplitude noise is smaller than the vacuum noise. 
In the large $p$ limit, the amplitude noise is half of the vacuum noise\cite{Ritsch92,Gheri92}. 

For two coupled NOPOs with $\langle \alpha_{s1}\rangle=\langle \alpha_{s2}\rangle=\sqrt{\frac{\gamma_p}{G}(p-1)}$, 
the normalized $HZ1$ criterion can be represented from small fluctuations of Wigner amplitudes:
\begin{equation}
\frac{HZ1}{\langle \hat{a}^{\dagger}_s\hat{a}_s\rangle}=1-2\langle |\Delta \alpha_{s1}|^2\rangle-2\langle \Delta \alpha_{s1}\Delta \alpha_{s2}\rangle. 
\end{equation}
These mean fluctuation products at a steady state can be calculated from 
\begin{equation}
\begin{bmatrix}
A' & A & -j & 0\\
A & A' & 0 & -j\\
-j & 0 & A' & A\\
0 & -j & A & A'
\end{bmatrix}
\begin{bmatrix}
\langle \Delta \alpha_{s1}^2\rangle \\
\langle |\Delta \alpha_{s1}|^{2} \rangle \\
\langle \Delta \alpha_{s1} \Delta \alpha_{s2} \rangle \\
\langle \Delta \alpha_{s1}^{*} \Delta \alpha_{s2} \rangle 
\end{bmatrix}
=\frac{1}{2}
\begin{bmatrix}
0 \\ 2+j \\0 \\ -j
\end{bmatrix}, 
\end{equation}
where $A=2(1-p^{-1})$ and $A'=A+j+\delta$. 
Here $\delta$ is a small value added to avoid divergence. 
The normalized $HZ1$ is finite even with $\delta \rightarrow 0$ and represented as 
\begin{equation}
\frac{HZ1}{\langle \hat{a}^{\dagger}_s\hat{a}_s\rangle}=\frac{j-2}{4j}-\frac{1}{4(p-1)}. 
\end{equation}
When $j=4$, this has a positive value for $p>3$. 

\section{Derivation and Numerical test of Eq.(\ref{fock2m})}

Here, we comment on the derivation and numerical simulation of Eq.(\ref{fock2m}). 
Derivation is similar to the diagonal case [Eq.(\ref{hyb1})], but we use the expansion with complex-$P$ representation for the pump mode. 
\begin{equation}
\label{hyb2}
\hat{\rho}=\sum_{N_s,N_s'} \int P_{N_s,N_s'}(\alpha_p,\alpha_p^{\dagger})\frac{|\alpha_p\rangle \langle \alpha_p^{\dagger *}|}{\langle \alpha_p^{\dagger *}|\alpha_p\rangle}\otimes |N_s\rangle \langle N_s'|d\alpha_pd\alpha_p^{\dagger}. 
\end{equation}
With this expansion, we can obtain an equation similar to Eq.(\ref{hybfp}): 
\begin{eqnarray}
\label{hybfp2}
\frac{\partial P_{N_s,N_s'}}{\partial t}=-\varepsilon \frac{\partial P_{N_s,N_s'}}{\partial \alpha_p}-\varepsilon \frac{\partial P_{N_s,N_s'}}{\partial \alpha_p^{\dagger}}\nonumber \\
+\frac{\partial}{\partial \alpha_p}(\gamma_p+G(1+N_s))\alpha_p P_{N_s,N_s'} \nonumber \\
+\frac{\partial}{\partial \alpha_p^{\dagger}}(\gamma_p+G(1+N_s'))\alpha_p^{\dagger} P_{N_s,N_s'} \nonumber \\
+\gamma_s[2\sqrt{(1+N_s)(1+N_s')}P_{N_s+1,N_s'+1} \nonumber \\
-(N_s+N_s')P_{N_s,N_s'}] \nonumber \\
+G\alpha_p^{\dagger}\alpha_p[2\sqrt{N_sN_s'} P_{N_s-1,N_s'-1}\nonumber \\
-((1+N_s)+(1+N_s'))P_{N_s,N_s'}].
\end{eqnarray}
The elimination of the pump mode is performed as $P_{N_s,N_s'}(\alpha_p,\alpha_p^{\dagger})=\rho_{N_s,N_s'}\delta \Bigl(\alpha_p - \frac{\varepsilon}{\gamma_p+G(1+N_s)}\Bigr)\delta \Bigl(\alpha_p^{\dagger} - \frac{\varepsilon}{\gamma_p+G(1+N_s')}\Bigr)$. 
Here, $\alpha_p$ depends on $N_s$ and $\alpha_p^{\dagger}$ depends on $N_s'$. 
Integrating with $\int d\alpha_pd\alpha_p^{\dagger}$, we can obtain Eq.(\ref{fock2m}). 

\begin{figure}[h]
\begin{center}
\includegraphics[width=8.0cm]{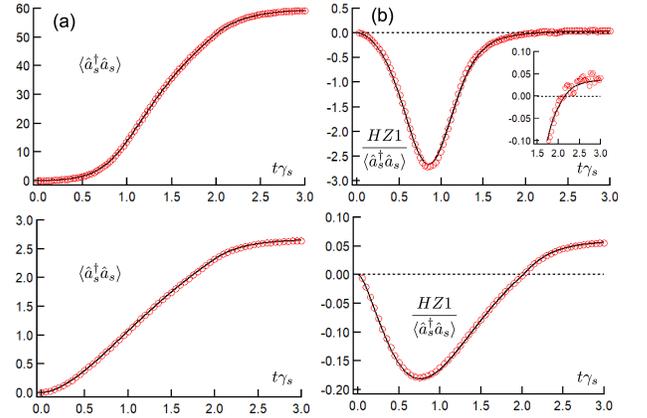}
\caption{Comparison between pump-eliminated Fock space approach and PSDE (upper) or 
direct calculation of Eq.(\ref{qmefull}) (lower) for two coupled NOPOs. 
Time-dependent (a) mean signal photon number $\langle \hat{a}_s^{\dagger}\hat{a}_s\rangle$, 
and (b) normalized $HZ1$ value. 
The results of pump-eliminated Fock space approach are shown by black lines and the results of the PSDE or 
the direct calculation are shown by red open circles. }
\label{app1}
\end{center}
\end{figure}

We can easily extend the pump-eliminated Fock space approach using Eq.(\ref{fock2m}) to two coupled NOPOs. 
First, we present the validation of the pump-eliminated Fock space approach by comparison with positive-$P$ SDE. 
We present the time development with $\gamma_p/\gamma_s=100$, $G/\gamma_s=5$, $J/\gamma_s=4$ and $p=4$. 
Excitation depended on time as $p\min(\sqrt{t\gamma_s/2},1)$. 
In the positive-$P$ calculation, we used Eqs.(\ref{pp1})-(\ref{pp4}) with an explicit pump mode. 
We consider $\Delta t\gamma_s=10^{-5}$ and the number of trajectories to be $9\times 10^{4}$. 
There were no trajectories with the divergence problem\cite{Gilchrist97}. 
For pump-eliminated Fock space calculation, we consider $\Delta t\gamma_s=2 \times 10^{-4}$ and 
the maximum photon number to be $N_s=100$. 
The deviation of the trace due to the cutoff of the Fock space was smaller than $10^{-6}$. 
In Fig.\ref{app1}(upper), we present the time development of the signal photon number and the normalized $HZ1$ criterion. 
As shown in the inset, $HZ1$ becomes positive and the entanglement criterion is satisfied. 
Next, in Fig.\ref{app1}(lower), we present the validation of the pump-eliminated Fock space approach 
by comparison with the direct calculation of Eq.(\ref{qmefull}). 
We present the time development with $\gamma_p/\gamma_s=50$, $G/\gamma_s=400$, $J/\gamma_s=9$ and $p=25$. 
In the direct calculation, $\Delta t \gamma_s=2\times 10^{-5}$, 
and the maximum pump (signal) photon number was $N_p=4$ ($N_s=13$). 

\end{document}